\documentclass[12pt]{article}
\usepackage[margin=1.9cm]{geometry}
\usepackage{graphicx}
\usepackage{cite}

\newcommand{\mysection}{\setcounter{equation}{0}\section}

\def\beq{\begin{equation}}
\def\eeq{\end{equation}}
\def\beqa{\begin{eqnarray}}
\def\eeqa{\end{eqnarray}}
 
\begin{document}

\begin{center}
{\Large \bf Top-quark production cross sections \\ at higher perturbative orders}
\end{center}

\vspace{2mm}

\begin{center}
{\large Nikolaos Kidonakis \let\thefootnote\relax\footnote{Presented at ICNFP 2025, Kolymbari, Crete, Greece}}\\
\vspace{2mm}
{\it Department of Physics, Kennesaw State University, \\
Kennesaw, GA 30144, USA}

\end{center}
 
\begin{abstract}
I present calculations of higher-order corrections to top-quark production via three different processes: the production of a top-antitop pair and a $Z$ boson; single-top quark production in the $s$-channel; and top-antitop production with a Higgs boson. It is shown that the contributions from soft-gluon corrections are numerically dominant and large in all these processes. I present approximate NNLO (aNNLO) and approximate N$^3$LO (aN$^3$LO) cross sections that include soft-gluon corrections added to the exact NLO QCD results. Electroweak corrections through NLO are also included for $t{\bar t}Z$ and $t{\bar t}H$ production. The theoretical predictions are compared to LHC measurements of total cross sections for these processes.
\end{abstract}
 
\mysection{Introduction}

The top quark is the heaviest elementary particle and, arguably, the most interesting. Its large mass and unique properties have made it a central topic of study at the LHC where numerous production modes have been observed and studied. Theoretical calculations for top quark cross sctions and differential distributions are at an advanced stage.
 
In this presentation, I discuss three different top-quark production processes: $t{\bar t}Z$ production, $s$-channel single-top production, and $t{\bar t}H$ production. All three processes are important for further study of the Standard Model and the search for physics beyond it. The $t{\bar t}Z$ process is important for measuring the coupling of the top quark to the $Z$ boson. Single-top production in the $s$-channel offers the possibility of measuring the $V_{tb}$ CKM matrix element. The $t{\bar t}H$ process, which involves the two heaviest Standard Model elementary particles, allows for a measurement of the Yukawa coupling of the top quark.

The perturbative corrections to these processes significantly increase the leading-order (LO) cross sections. Furthermore, one can calculate certain contributions beyond next-to-leading order (NLO) or next-to-next-to-leading order (NNLO) that arise from soft-gluon resummation and which are numerically dominant. Thus, approximate next-to-next-to-next-to-leading order (aN$^3$LO) theoretical predictions for cross sections can be derived and compared to recent experimental data from the LHC for all three of the processes studied here.

In the next section, I briefly discuss the formalism for the calculation of soft-gluon corrections in top-quark processes. Section 3 has results for total cross sections and top-quark differential distributions in $t{\bar t}Z$ production at LHC energies through aN$^3$LO QCD and also with the inclusion of NLO electroweak (EW) corrections. In Sec.~4, results are presented for the total cross section for $s$-channel single-top and single-antitop production at LHC energies through aN$^3$LO QCD. In Sec.~5, I present results for the $t{\bar t}H$ cross section at the LHC through aN$^3$LO QCD + NLO EW. The Conclusions are given in Sec.~6.

\mysection{Top-quark production and soft-gluon corrections}

The formalism for soft-gluon resummation in QCD hard-scattering processes has been developed in several works \cite{NKGS,LOS,NKtt,NKsingletop,NKtt2l,FK2020}. The soft-gluon corrections are theoretically and numerically important for top-quark processes, and they approximate known exact results for total and differential cross sections at NLO and NNLO very well. 

We study partonic processes of the form $a(p_a)\, + \, b\, (p_b) \to t(p_t)\, + \, X$, with 
kinematical variables $s=(p_a+p_b)^2$, $t=(p_a-p_t)^2$, $u=(p_b-p_t)^2$, and  
we define the threshold variable  $s_4=s+t+u-m_t^2-m_X^2$. At partonic threshold $s_4 \to 0$.
The soft-gluon contributions are of the form $\left[\frac{\ln^k(s_4/m_t^2)}{s_4}\right]_+$ with $k \le 2n-1$
in the order $\alpha_s^n$ corrections. We resum these soft corrections for the double-differential cross section in one-particle-inclusive (1PI) kinematics. 

Soft-gluon resummation is derived using Laplace (1PI) or Mellin transforms, since then the cross section is factorized into functions that describe soft and collinear emission from the colliding partons, a hard function $H_{ab \to tX}$ for the hard scattering, and a soft function $S_{ab \to tX}$ for noncollinear soft-gluon emission \cite{NKGS}. Renormalization group evolution results in the expression for the resummed cross section:
\beqa
&& {\hat{\sigma}}_{ab \to tX}^{res}(N) =   
\exp\left[ \sum_{i=a,b} E_i(N_i,\mu_F)\right] \; 
\exp\left[ E'(N)\right] \; {\rm tr} \left \{H_{ab \to tX} \left(\alpha_s(\sqrt{s})\right) \right. 
\nonumber\\ && \hspace{-16mm} \left. \times \,
{\bar P} \exp \left[\int_{\sqrt{s}}^{{\sqrt{s}}/{\tilde N}} \frac{d\mu}{\mu} \Gamma^{\dagger}_{S \, ab \to tX} \left(\alpha_s(\mu)\right)\right] \,
S_{ab \to tX} \left(\alpha_s(\sqrt{s}/{\tilde N}) \right) \, P \exp \left[\int_{\sqrt{s}}^{{\sqrt{s}}/{\tilde N}} \frac{d\mu}{\mu}\; \Gamma_{S \, ab \to tX}
\left(\alpha_s(\mu)\right)\right] \right\}
\label{rgeres}
\eeqa
where the first two exponents resum universal contributions from incoming and outgoing partons, and the soft function evolves via the soft anomalous dimensions $\Gamma_{S \, ab \to tX}$, which are known at two loops or higher \cite{NKGS,NKsingletop,NKtt2l,FK2020,NK2loop,NKst3l,NK4loop}.

Finite-order expansions in physical space can be derived with no prescription needed or used. By expanding to third order, and matching with known exact fixed-order results at lower orders, we derive aN$^3$LO predictions for cross sections and distributions.

\mysection{$t{\bar t}Z$ production}

Measurements of $t{\bar t}Z$ cross sections have been performed at 7 TeV \cite{CMS7ttZ}, 8 TeV \cite{CMS8ttZ,ATLAS8ttZ}, and 13 TeV \cite{ATLAS13ttZ,CMS13ttZ} collisions by CMS and ATLAS at the LHC. The QCD corrections at NLO \cite{LMMP,KTP} are large, around 32\% for collisions at 13.6 TeV energy. The electroweak corrections \cite{FHPSZ} are smaller but significant. Further improvement in theoretical accuracy may be achieved by the inclusion of higher-order soft-gluon corrections \cite{KFttZ}. 

The NLO expansion of the resummed cross section closely approximates numerically the exact NLO results for the total cross section and the top-quark $p_T$ and rapidity distributions. We calculate second-order and third-order soft-gluon corrections and also include EW corrections at NLO, thus deriving aN$^3$LO QCD + NLO EW cross sections for $t{\bar t}Z$ production \cite{KFttZ} which are the state-of-the-art.

\begin{table}[htb]
\begin{center}
\begin{tabular}{|c|c|c|c|c|c|} \hline
\multicolumn{6}{|c|}{$t{\bar t} Z$ cross sections in $pp$ collisions at the LHC} \\ \hline
$\sigma$ in fb & 7 TeV & 8 TeV & 13 TeV & 13.6 TeV & 14 TeV \\ \hline
aNNLO QCD             & $163^{+7}_{-10}$ & $245^{+10}_{-15}$ & $952^{+29}_{-48}$ & $1074^{+33}_{-54}$ & $1157^{+35}_{-58}$ \\ \hline
aN$^3$LO QCD          & $168^{+5}_{-8}$ & $253^{+8}_{-12}$ & $982^{+25}_{-28}$ & $1108^{+28}_{-32}$ & $1194^{+30}_{-34}$ \\ \hline
aN$^3$LO QCD + NLO EW & $173^{+5}_{-8}$ & $261^{+7}_{-12}$ & $998^{+21}_{-26}$ & $1125^{+24}_{-30}$ & $1211^{+25}_{-30}$ \\ \hline
\end{tabular}
\caption[]{The $t{\bar t}Z$ cross sections \cite{KFttZ} with scale uncertainties at orders from aNNLO QCD through aN$^3$LO QCD + NLO EW, in $pp$ collisions at the LHC with $\sqrt{S}=7$, 8, 13, 13.6, and 14 TeV, $m_t=172.5$ GeV, and MSHT20 NNLO pdf.}
\label{tablettZ}
\end{center}
\end{table}

In Table \ref{tablettZ}, we show results \cite{KFttZ} for the total cross section with scale uncertainties at various LHC energies using MSHT20 NNLO pdf \cite{MSHT20NNLO}. The central results are with a scale $\mu=m_t=172.5$ GeV. At 13.6 TeV, the NLO QCD corrections provide an enhancement of 32\%, the aNNLO QCD corrections add 12\%, the aN$^3$LO QCD corrections add a further 5\%, and the electroweak NLO corrections are 2\%. Thus, the total aN$^3$LO QCD+NLO EW cross section is 52\% bigger than that at LO QCD.

Next, we make a comparison with 7, 8, and 13 TeV LHC data.
At 7 TeV, the measurement from CMS is $0.28^{+0.14}_{-0.11}{}^{+0.06}_{-0.03}$ pb \cite{CMS7ttZ}. 
At 8 TeV, there are measurements from ATLAS of $176^{+58}_{-52}$ fb \cite{ATLAS8ttZ} and from CMS of $242^{+65}_{-55}$ fb \cite{CMS8ttZ}.  At 13 TeV, there are measurements of $0.95 \pm 0.05 \pm 0.06$ pb from CMS \cite{CMS13ttZ},
and of $0.99 \pm 0.05 \pm 0.08$ pb and $0.86 \pm 0.04 \pm 0.04$ pb from ATLAS \cite{ATLAS13ttZ}. 
The data are in agreement with the theoretical predictions.

With MSHT20 aN$^3$LO pdf \cite{MSHT20aN3LO}, we find aN$^3$LO QCD + NLO EW theoretical cross sections of $170^{+5}_{-8}$ fb at 7 TeV, $255^{+7}_{-12}$ fb at 8 TeV, $974^{+20}_{-25}$ fb at 13 TeV, $1096^{+23}_{-29}$ fb at 13.6 TeV, and $1182^{+24}_{-29}$ fb at 14 TeV.

\begin{figure}[htb]
\begin{center}
\includegraphics[width=88mm]{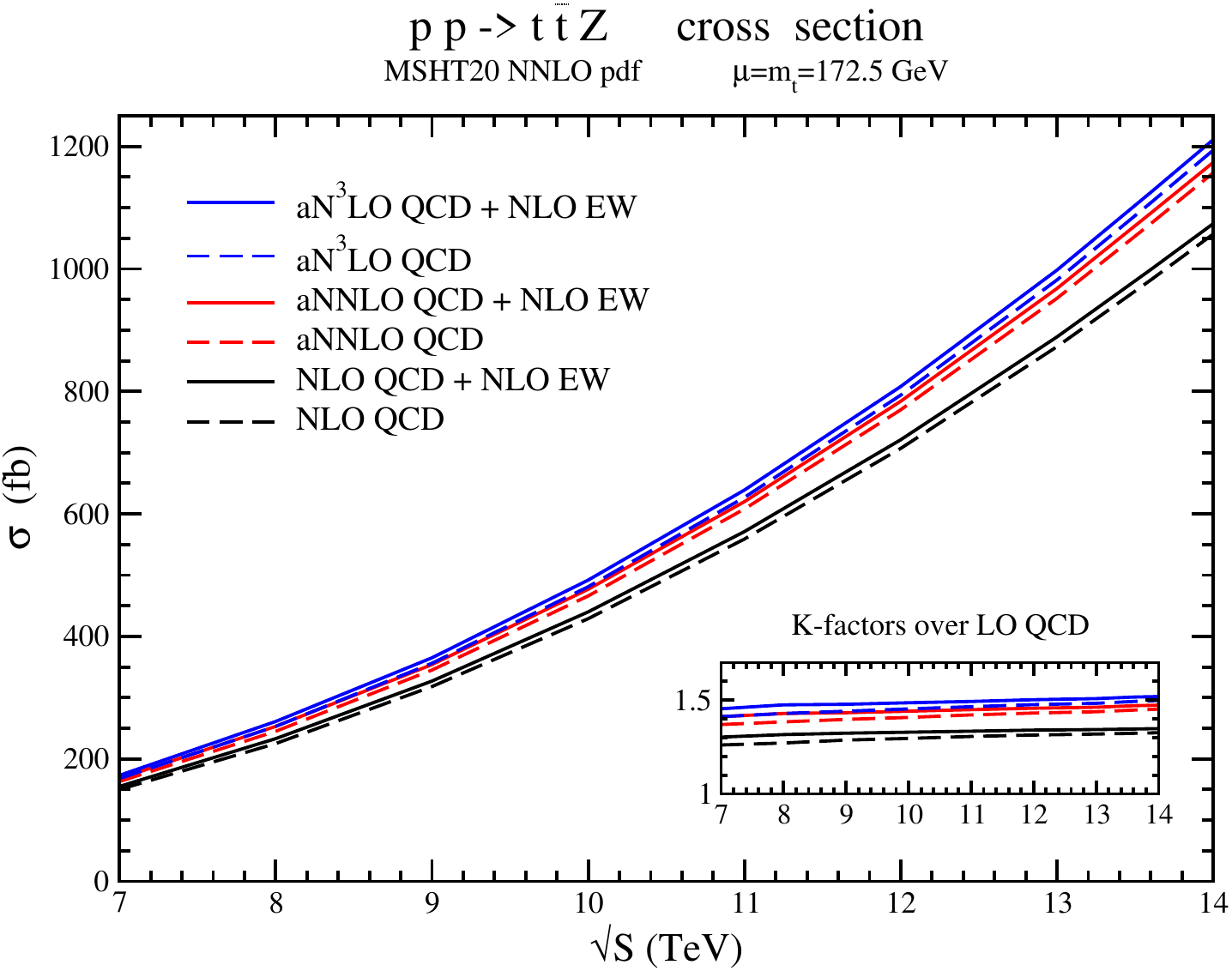}
\includegraphics[width=88mm]{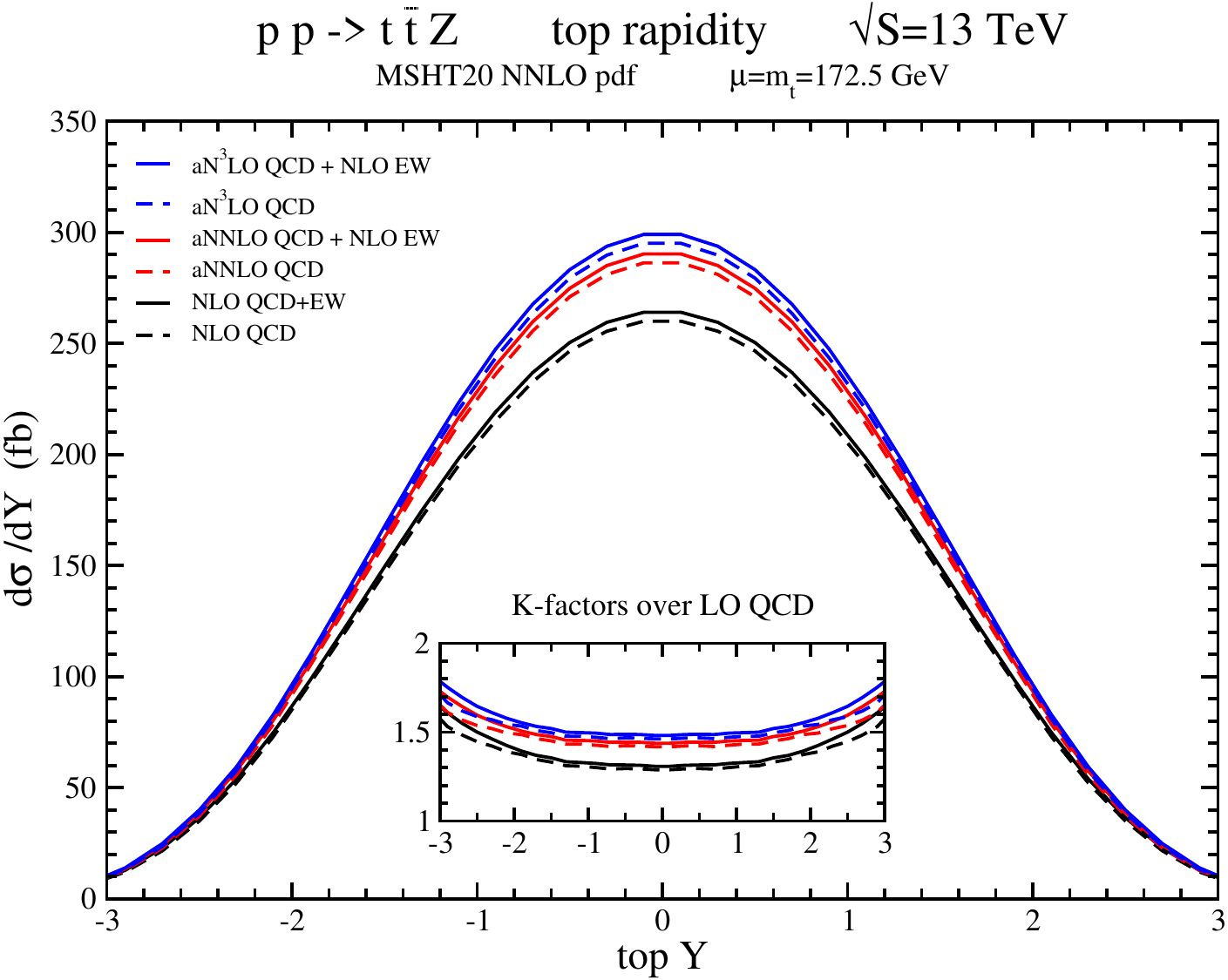}
\caption{The total cross sections (left) and top-quark rapidity distributions (right) for $t{\bar t}Z$ production in $pp$ collisions at LHC energies. Results are shown through aN$^3$LO QCD + NLO EW, and the inset plots display the $K$-factors relative to LO QCD.} 
\label{ttZcsy}
\end{center}
\end{figure}

Fig. \ref{ttZcsy} shows the total cross section for a range of LHC energies (left plot) and the top-quark rapidity distribution at 13 TeV energy (right plot) for $t{\bar t}Z$ production. Results are shown at NLO, aNNLO, and aN$^3$LO QCD with and without the addition of NLO EW corrections. The $K$-factors are large in all cases. In the rapidity distribution, we note that the $K$-factors increase significantly at larger rapidities.

\mysection{$s$-channel single-top production}

Single-top production in the $s$-channel has been actively studied at the LHC, and several cross section measurements have been taken \cite{ATLAS8sch,CMS7and8sch,ATLASCMSsch,ATLAS13sch}.

NNLO results (where color exchanges between heavy-quark and light-quark lines are neglected) for the total cross section have been presented in Ref. \cite{NNLOsch}.
Further improvement in theoretical accuracy may be achieved by the inclusion of higher-order soft-gluon corrections. 
Resummation for $s$-channel single-top production has been presented in Refs. \cite{NKsingletop,NKst3l,NKstLHC,NKsch,NKstdiff,NKs25}.
In the latest work \cite{NKs25}, the two-loop soft anomalous dimensions were calculated additionally for a massive treatment of final-state $b$-quarks. However, we set $m_b=0$ for the numerical calculations.
The NNLO expansion of the resummed cross section closely approximates numerically the NNLO result for the total cross section in Ref. \cite{NNLOsch}, within a few per mille for the central value and with very similar scale uncertainty.
The aN$^3$LO QCD theoretical prediction \cite{NKs25} is the state of the art.

\begin{table}[htb]
\begin{center}
\begin{tabular}{|c|c|c|c|c|c|} \hline
\multicolumn{6}{|c|}{Single-top plus single-antitop $s$-channel cross sections in $pp$ collisions} \\ \hline
$\sigma$ in pb & 7 TeV & 8 TeV & 13 TeV & 13.6 TeV & 14 TeV \\ \hline
NNLO QCD & $4.67^{+0.05}_{-0.04}$ & $5.72^{+0.06}_{-0.05}$ & $11.2^{+0.08}_{-0.06}$ & $11.9^{+0.09}_{-0.06}$ & $12.4^{+0.09}_{-0.06}$  \\ \hline 
aN$^3$LO QCD & $4.79^{+0.04}_{-0.04}$ & $5.86^{+0.04}_{-0.05}$ & $11.5^{+0.05}_{-0.06}$ & $12.2^{+0.06}_{-0.06}$ & $12.6^{+0.07}_{-0.06}$  \\ \hline
\end{tabular}
\caption[]{The single-top plus single-antitop $s$-channel cross sections \cite{NKs25} with scale uncertainties through aN$^3$LO in $pp$ collisions at the LHC for various values of $\sqrt{S}$, with $m_t=172.5$ GeV and MSHT20 aN$^3$LO pdf.}
\label{tablesch}
\end{center}
\end{table}

In Table \ref{tablesch}, we present the total cross sections with scale uncertainties for single-top plus single-antitop $s$-channel production at LHC energies. The central results are with $\mu=m_t=172.5$ GeV and MSHT20 aN$^3$LO pdf \cite{MSHT20aN3LO}.

At 13.6 TeV,
the NLO QCD corrections are 35\%,
the NNLO QCD corrections are an additional 7\%,
and the aN$^3$LO QCD corrections are a further  3\%.
Thus, the total aN$^3$LO cross section is 46\% bigger than the result at LO QCD.

\begin{figure}[htb]
\begin{center}
\includegraphics[width=10cm]{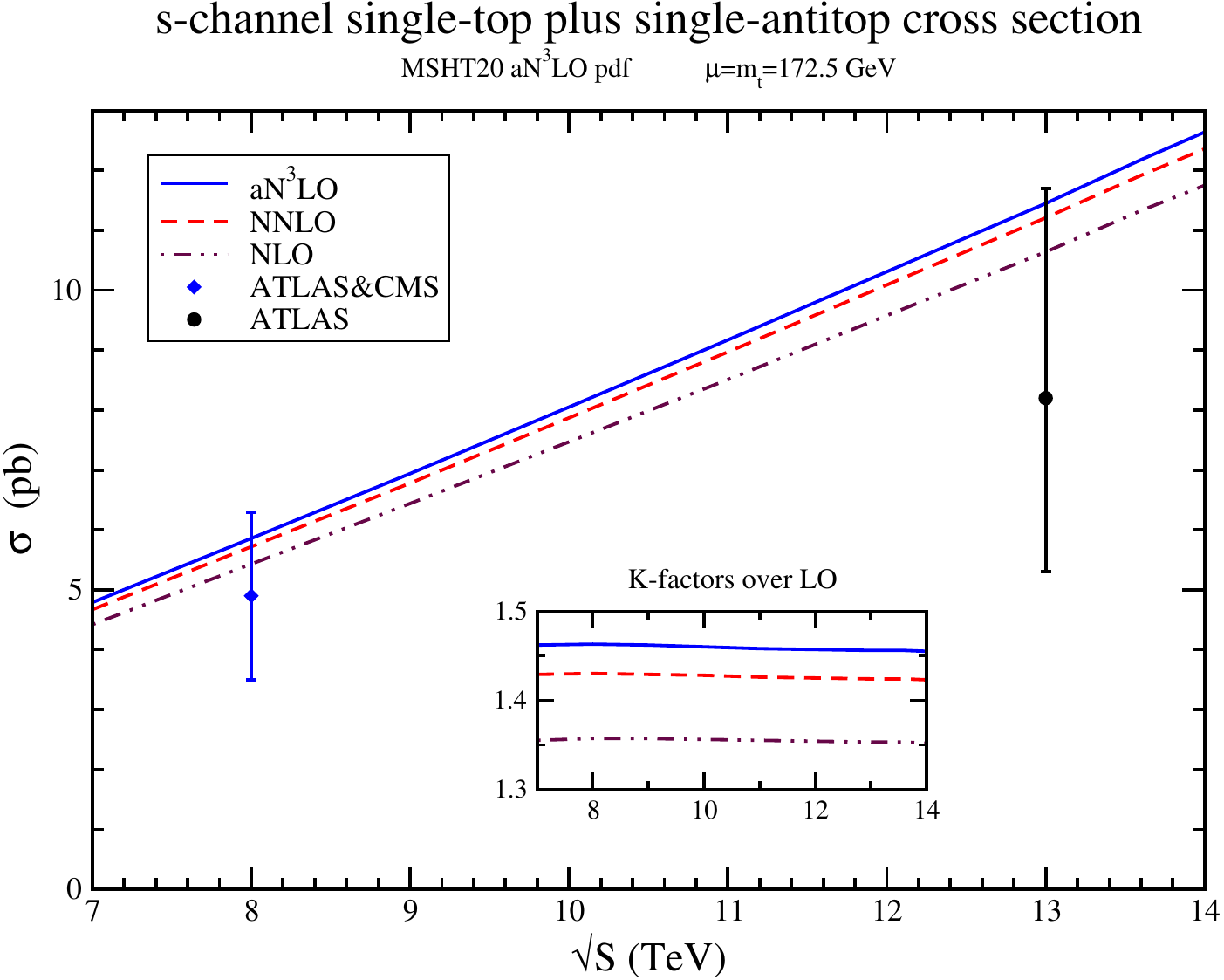}
\caption{The total cross sections for $s$-channel single-top plus single-antitop production through aN$^3$LO \cite{NKs25} in $pp$ collisions at LHC energies. Also shown are data at 8 TeV \cite{ATLASCMSsch} and 13 TeV \cite{ATLAS13sch} LHC energies. The inset plot displays the $K$-factors relative to LO.}
\label{schannelplot}
\end{center}
\end{figure}

In Fig. \ref{schannelplot}, we plot the total $s$-channel cross section for a range of LHC energies and compare with recent data. The current experimental error bars indicate an uncertainty that is much bigger than the theoretical one.
At 8 TeV, we find an aN$^3$LO cross section (with scale and pdf uncertainties) of $5.86^{+0.04}_{-0.05}{}^{+0.11}_{-0.14}$ pb \cite{NKs25} 
compared to the ATLAS+CMS measurement of $4.9 \pm 1.4$ pb \cite{ATLASCMSsch}.
At 13 TeV, we find an aN$^3$LO cross section of $11.5^{+0.05}_{-0.06}{}^{+0.13}_{-0.29}$ pb \cite{NKs25}
compared to the ATLAS measurement of $8.2^{+3.5}_{-2.9}$ pb \cite{ATLAS13sch}.

\mysection{$t{\bar t}H$ production}

The associated production of a top-quark pair with a Higgs boson is a process that has been the subject of intense study at LHC energies. The ATLAS and CMS experiments have made multiple measurements of $t{\bar t}H$ cross sections at LHC energies \cite{ATLASttH,CMSttH}.

Results for $t{\bar t}H$ production at partial NNLO QCD accuracy have been presented in Refs. \cite{NNLO1ttH,NNLO2ttH}, and results also including NLO EW corrections and soft-gluon resummation in other formalisms and kinematics was given in Ref. \cite{NNLOresttH}. In Ref. \cite{NKNYttH}, aNNLO and aN$^3$LO corrections were calculated in 1PI kinematics to make state-of-the-art predictions for the total $t{\bar t}H$ cross section as well as the top-quark $p_T$ and rapidity distributions.

\begin{table}[htb]
\begin{center}
\begin{tabular}{|c|c|c|c|} \hline
\multicolumn{4}{|c|}{$t{\bar t}H$ cross sections in $pp$ collisions at the LHC} \\ \hline
$\sigma$ in fb & 13 TeV & 13.6 TeV & 14 TeV \\ \hline
aNNLO QCD & $518^{+4}_{-14}{}^{+12}_{-13}$ & $581^{+4}_{-15}{}^{+13}_{-14}$ & $626^{+5}_{-16}{}^{+14}_{-15}$ \\ \hline
aN$^3$LO QCD & $523^{+3}_{-7}{}^{+12}_{-13}$ & $587^{+3}_{-7}{}^{+13}_{-14}$ & $633^{+3}_{-8}{}^{+14}_{-15}$ \\ \hline
aN$^3$LO QCD + NLO EW & $529^{+3}_{-6}{}^{+12}_{-13}$ & $593^{+3}_{-6}{}^{+13}_{-14}$ & $639^{+3}_{-7}{}^{+14}_{-15}$ \\ \hline
\end{tabular}
\caption[]{The $t{\bar t}H$ cross sections \cite{NKNYttH} with QCD and EW corrections through aN$^3$LO QCD + NLO EW, with scale and pdf uncertainties, in $pp$ collisions at the LHC with $\sqrt{S}=13$, 13.6, and 14 TeV, $m_t=172.5$ GeV, $m_H=125.2$ GeV, and MSHT20 aN$^3$LO pdf.}
\label{tablettH}
\end{center}
\end{table}

In Table \ref{tablettH}, we show results with scale and pdf uncertainties for the total $t{\bar t}H$ production cross section \cite{NKNYttH} at LHC energies. The central results are with $\mu=m_t=172.5$ GeV, $m_H=125.2$ GeV, and MSHT20 aN$^3$LO pdf \cite{MSHT20aN3LO}. We observe a big reduction in scale uncertainty at higher orders.

At 13.6 TeV LHC energy, the NLO QCD corrections are 15\%, the aNNLO QCD corrections are 5\%, the aN$^3$LO QCD corrections are 1\%, and the NLO EW corrections are 1\%. Thus, the sum of the corrections provides a 22\% enhancement over LO QCD.

\mysection{Conclusions}

I have presented higher-order corrections for three top-quark production processes: $t{\bar t}Z$ production, $s$-channel single-top production, and $t{\bar t}H$ production. Soft-gluon resummation was used to calculate aNNLO and aN$^3$LO soft-gluon corrections.

For $t{\bar t}Z$ production, calculations were presented through aN$^3$LO QCD + NLO EW for the total cross section and the top-quark rapidity distribution. The higher-order perturbative corrections increased the cross section by 52\% over LO QCD at 13.6 TeV LHC energy. The enhancements were even larger for the rapidity distribution at large top-quark rapidity values.

For $s$-channel single-top production, there is excellent agreement of aNNLO with NNLO QCD results. The total QCD corrections through aN$^3$LO are 46\% bigger than LO QCD in proton collisions at 13.6 TeV energy.

For $t{\bar t}H$ production, the aN$^3$LO QCD + NLO EW predictions are 22\% higher than LO QCD for 13.6 TeV proton collisions. In all cases, the higher-order corrections further enhance and improve the theoretical predictions via a reduction in scale dependence.

\mysection*{Acknowledgements}
This material is based upon work supported by the National Science Foundation under Grant No. PHY 2412071.

\end{document}